\documentclass[aps,prb,showpacs,twocolumn,amsmath,amssymb,superscriptaddress,floatfix]{revtex4}

\usepackage{epsfig}
\usepackage{epstopdf}
\usepackage{graphicx}
\usepackage{grffile}
\usepackage{bm}
\usepackage{sidecap}
\usepackage{mathptmx}
\usepackage{txfonts}
\usepackage[colorlinks,citecolor=blue,linkcolor=Fuchsia]{hyperref}
\usepackage[usenames,dvipsnames,svgnames]{xcolor}
\usepackage{array}
\usepackage{float}

\usepackage{lipsum}

\newcommand{\tobedeleted}[1]{\textcolor{green}{#1}}
\renewcommand{\tobedeleted}[1]{\relax}

\begin{document}
\renewcommand{\thefigure}{\arabic{figure}}
\def\be{\begin{equation}}
\def\ee{\end{equation}}
\def\ber{\begin{eqnarray}}
\def\eer{\end{eqnarray}}

\title{Effect of Electron-RBM Phonon Interaction on Conductance of Metallic Zigzag Carbon Nanotubes}
\date{\today}

\author {Reyhaneh Taj}\email{reyhanetadj@gmail.com}
\affiliation{Department of Physics, Iran university of Science and Technology, Narmak, Tehran, Iran}

\author {Afshin Namiranian}
\affiliation{Department of Physics, Iran university of Science and Technology, Narmak, Tehran, Iran}

\begin{abstract}
We use the energy analysis as a perturbative method to study the effect of electron-radial breathing mode (RBM) phonon interaction on the electrical conductance of long metallic zigzag carbon nanotubes (CNTs). The band structure of zigzag CNTs is calculated by exerting zone-folding method on relations derived by using the nearest neighbor approximation of tight-binding expression for the $\pi$ bands of graphene. The small hollow cylinder model, with two different approximations, is used to obtain the RBM frequency in our calculation. As the result, we have calculated the effects of electron$ - $RBM phonon interaction on the conductance of zigzag CNTs. It has been observed that current is a step$ - $like function of bias voltage due to the absorption or emission by electron injection in the system. Moreover, the dependence of the conductance to the temperature in low bias and high bias voltages has been studied. In this paper, we propose a simple and useful method for phonon spectroscopy. Also, since RBM mode determines the geometry and structure of CNT, this approach can be used for characterization of CNTs.
\end{abstract}
\maketitle

\section{Introduction}
The appearance of carbon nanotubes\cite{white-nat393, mc-prl83} (CNTs) as promising building blocks for electron flow have established a novel revolution in science and technology of nano$ - $scaled devices. Since in comparison with other systems, scattering factors have less effect on CNT's conductance, it becomes essential to perceive the electron transport properties of these quasi$ - $one$ - $dimensional systems\cite{JDMPP}. CNTs can be considered as a rolled graphene sheet which can be in the form of single$ - $wall or multi$ - $wall. Moreover, single$ - $wall CNTs (SWCNTs) can be categorized into two forms of chiral and achiral tubes that the achiral ones are divided into zigzag and armchair. While armchair CNTs are metals, zigzag CNTs can be metal or semiconductor based on their geometry. 
In practice, in the production of CNTs, they would be produced with several diameter sizes together, so it would be important to propose a method to detect the CNT's diameter size in the system. 

CNTs demonstrate diverse significant properties\cite{JCXS, ANBGFG, MPMD, FAJM, AMSMEAP}, importantly, one of these remarkable properties is elevated conductance in the ballistic regime\cite{white-nat393}, as it has been shown that the current density of CNTs can reach near $10^{9}\frac{cm}{\AA}$\cite{PJRPSR}. This, together with being quasi-one-dimensional system make CNTs a novel candidate to be used as metallic interconnecting systems (quantum wires)\cite{FAPGSWMW, JMYSZPMH, HMSRDEMSTM, HFiedler}. 

Beside their application as conducting wires\cite{TMHHARELJC}, CNTs can be used as field$ - $effect transistors\cite{MDHPLngAArE}, or single$ - $electron$ - $tunnelling transistors\cite{SJRMCNATUR}. 
Although, CNTs are good ballistic transporter one can still consider several proposals for scattering in these materials. One of the main mechanisms is the electron$ - $phonon (e$ - $ph) interaction which is an important scattering mechanism in the CNTs for a wide range of temperatures.

Novel properties and one$ - $dimensional structure of CNTs yield to emerge unique modes and properties in their phonon spectrum \cite{SCJ}. Characterisation of CNTs has been done extensively by Raman spectroscopy \cite{MAASMEBAPRG, ZLJPC, ARJCMTGM, SDPMM, MSGRAPR} , and Dependant on the type of vibrations of carbon atoms in a long CNT, three types of phonon modes exist: $ 1) $ RBM phonon mode that comes from coherent radial vibrations of carbon atoms; $ 2) $ Longitudinal phonon modes that are due to vibrations of carbon atoms in the direction of longitudinal axis of tube; $ 3) $ Transverse phonon modes that model vibrations of carbon atoms perpendicular to the longitudinal axis of the tube. Although, both transverse and longitudinal phonon modes have two optical and acoustical branches, the RBM phonon mode is just optical. Raman scattering spectra show three peaks in the energy, the G-band, D-band, and radial breathing mode (RBM) which can be used as the distinctive characteristic of SWCNTs \cite{AMERSBCPCKASKR}. RBM peaks, which appear in the lower frequency region $(<400 cm^{ −1})$, are used to establish the tube diameter ($d$) and chirality ($n_1,n_2$) analysis based on the resonant Raman scattering effects\cite{SKDJNATY, KRGMSHJ}. Then RBM describes nanotube uniquely \cite{RBPRBVNTK} and it is used for characterization of the nanotube in laboratory \cite{HDJC}.
\\
Phonons affect strongly CNTs conductance, and e$ -  $ph coupling plays a crucial part in the perception of properties of CNTs. 
In low-bias regime, because of electron-acoustic phonon interaction’s weakness, it is observed that, ballistic conductance occurs\cite{HT, LG, GN}. In high-bias regime, high-energy vibrational modes are excited and interaction between electrons and these phonons restricts ballistic conductance \cite{SJFR}. e-ph interaction in metallic CNTs has a strong effect on the current behavior in different temperatures\cite{CERF}.
\\
In this paper, we study the effect of electron$ - $RBM phonon interaction in metallic zigzag CNT on variations of current by using energy analysis approach. It has been observed that due to the electron$ - $RBM phonon interaction, the current changes in a step$ - $like form as a function of bias voltage. It means that creating a new phonon mode, in addition to the others, gives rise to changing the current for certain values of energy changing. Temperature is a significant parameter in the conductance of CNTs\cite{TDERJ} as our results have demonstrated that in low temperature and bias voltage, e$ -  $ph interaction is not an important factor in current changes; but in higher temperatures because of excitation of high energy optical phonon modes, the e$ -  $ph interaction has more important effect on the decreasing of the current.

The arrangement of this paper is as follows. First, in section \ref{EDR}, the electron dispersion relation for zigzag CNTs is presented. In Sec. \ref{RBM}, we present both approximations of the RBM frequency as well as Fr\"ohlich Hamiltonian to describe the e$ -  $ph coupling with the corresponding coefficient for this type of atom vibrations. Further, to tackle the signature of interaction between the electron and RBM phonon on the correction of the current, we consider a CNT coupled to two reservoirs and apply energy analysis method in Sec. \ref{cond}. We focus on the obtained results in the Sec. \ref{results}, and describe how the temperature and the voltage affect the current. Finally, the paper is briefly concluded and summarized in Sec. \ref{conclusion}.

\section{Theory}
\subsection{Electronic structure of CNTs}\label{EDR}
The structure of a SWCNT is uniquely defined by the chiral vector $\textbf{C}$ which indicates the rolling up direction. 
Since CNT is a cylindrically rolled counterpart of a graphene sheet, $\textbf{C}$ in terms of unit vectors of graphene, is expressed as, $\textbf{C}=n_{1}\textbf{a}_{1}+n_{2}\textbf{a}_{2}$ in which $\textbf{a}_{1}$ and $\textbf{a}_{2}$ are chiral indexes. Besides, to calculate the electron dispersion relation of zigzag CNTs, we apply zone-folding method \cite{SDDPh} on the relation derived by the tight binding model of graphene under nearest neighbor approximation\cite{SCJ} for $\pi$-electrons. In this method, by using periodic boundary conditions in the circumferential direction denoted by the chiral vector $C$, the wave vector associated with the $C$ direction becomes quantized. Thus, the energy bands consist of a set of one-dimensional energy dispersion relations. Assuming the vanishing orbital overlap, electron dispersion relation of $(n,0)$ zigzag CNTs would be
\begin{equation}\label{Ek}
E^{\pm}_{zz}(m,k_{z})=\gamma_0 \; \sqrt{3+2 \cos (\frac{2\pi m}{n})+4 \; \cos (\frac{\pi m}{n}) \cos (\pi k_z)}.
\end{equation}
Here, $\gamma_0\approx 3.033$eV, is the nearest neighbor hopping energy, $k_z$ shows the component of the wave vector parallel to the CNT's axis where $m$ and $n$ satisfy $-(n-1) \leq m \leq n $. For a general $(n,0)$ zigzag carbon nanotubes, when $n$ is a multiple of $3$, the energy gap becomes zero at $k=0$ and CNTs show metallic behavior.

\subsection{Coupling between electron and Radial-breathing mode phonon}\label{RBM}
e$ -  $ph interaction has a key role in the perception of electronic, optical and transport properties of CNTs \cite{ST}. To address CNTs identification in particular of their chirality, we study the effect of radial-breathing mode of phonons on CNTs conductance. Coherent vibrations of carbon atoms in the direction of nanotube diameter result in RBM phonons and their frequency depends on the inverse of CNTs diameter \cite{HDJC, AMERSBCPCKASKR, SJJDJBMASJY}. In this paper we considered two approximations for the frequency of RBM mode. In the first one, the frequency of RBM phonons can be achieved from the continuum mechanics of a small hollow cylinder shown by Mahan \cite{GDMAHAN, MCMRcardona} as
\begin{equation}
\omega_{RBM}=\frac{c_1}{d}+c_2
\end{equation}
where for an isolated nanotube, theoretical and experimental reported values of constant coefficients are $c_1=218$ to $248 \, cm^{-1}nm$ and $c_2=0$ \cite{JPG, SCJ}. By adopting the approximation used in Ref. \cite{HDJC}, for metallic CNTs, we consider $c_1=243\,cm^{-1}nm$ and $c_2=0$. 
\\
For small diameter CNTs, a more precise approximtion can be used as a tensional force results in distortaion of its band structure\cite{JGH}. The $ \omega_{RBM} $ in this approximation (the second approximation) would read as
\begin{equation}\label{omega2}
\omega_{RBM}=\frac{c_1}{d}+\frac{c_2}{d^3},
\end{equation}
where $c_1=226 cm^{-1}$ nm and $c_2=1.5\pm0.5$ cm$^{-1}$ $ nm^3 $ \cite{HDJC}. For small diameter nanotubes, effects of rolling nanotubes cause noticeable deviation from the simple appropriation of $\omega_{RBM}$ with the inverse of CNT diameter. In order to model the e$ -  $ph interaction, Fr\"ohlich proposed the below Hamiltonian which is especially suitable for transport. Assuming that e$-$ph coupling occurs with the same coefficient, the e$-$ph interaction Hamiltonian can be written as
\begin{equation}\label{Hint}
H_{e-ph}=\sum_{k, k^\prime} M_{k, k^\prime} \,(a^{\dagger}_{-q}+a_{q}) \, c^{\dagger}_{k}c_{k^\prime},
\end{equation}
where $c^{\dagger}_{k}$ and $c_{k^\prime}$ ($a^{\dagger}_{-q}$ and $a_{q}$) are Fermionic (Bosonic) creation and annihilation operators respectively, $M_{k, k'}$ shows e$ -  $ph coupling coefficient and sum is over all electronic states. The diagonal matrix elements of the e$ -  $ph coupling Hamiltonian for optical phonons can be obtained from the shift of the electronic bands under deformation of the atomic structure corresponding to the phonon$ - $pattern \cite{MCMRcardona}
\begin{equation}
M_{e-ph}=\sqrt{\frac{\hbar}{2 m \, N \, \omega_{RBM}}} \sum_{a} \varepsilon_a \frac{\partial E_b(k)}{\partial u_a},
\end{equation}
in which $m$ is the atomic mass of electron, $N$ represents the number of unit cells and $a$ indixes the atoms in the unit cell of the nanotube. $\varepsilon_a$ refers to the normalized phonon eigenvector and $\frac{\partial E_b(k)}{\partial u_a}$ describes the changes in the electronic energy $E_b$ due to the atomic displacement $u_a$. Because $\frac{\partial E_b(k)}{\partial u_a}$ is proportional to $d^{-1}$ \cite{MSHJPC}, $M_{k, k^\prime}$ can be written as
\begin{equation}
M_{k, k^\prime}=\frac{1}{d}\sqrt{\frac{\hbar}{2 m \, N \, \omega_{RBM}}}.
\end{equation}

\subsection{conductance calculation}\label{cond}
Now consider a bias voltage $eV <<E_F$, where $E_F$ is the Fermi energy of electrons in the nanotube, applied on the mesoscopic structure including a metallic zigzag SWCNT, which its length is much larger than its diameter and it is connected to metallic reservoirs.  It is assumed that the length of two metallic electrodes is in the order of $\lambda_F$, hence, the transportation of ballistic electrons through the nanotube is robust against the edge effect \cite{AYA, AJ}. In addition, the rate reduction of the electric field in the nanotube is proportional to $a/L$ (where $1.25  < a < 1.75$ $\AA$) and $L$ is the nanotube length \cite{AM}. The conduction of the nanotube can be divided into two main parts: edge and central parts\cite{LS}. Also, it has been assumed that the electron and phonon population is in balance and they behave independently. Thus, the e$ -  $ph interaction can be considered as a perturbative phenomenon. In the following, we investigate the influence of e$ -  $ph coupling on the changes of current in the central part of a nanotube.
\begin{figure}[!h]
  \includegraphics[width=0.48\textwidth]{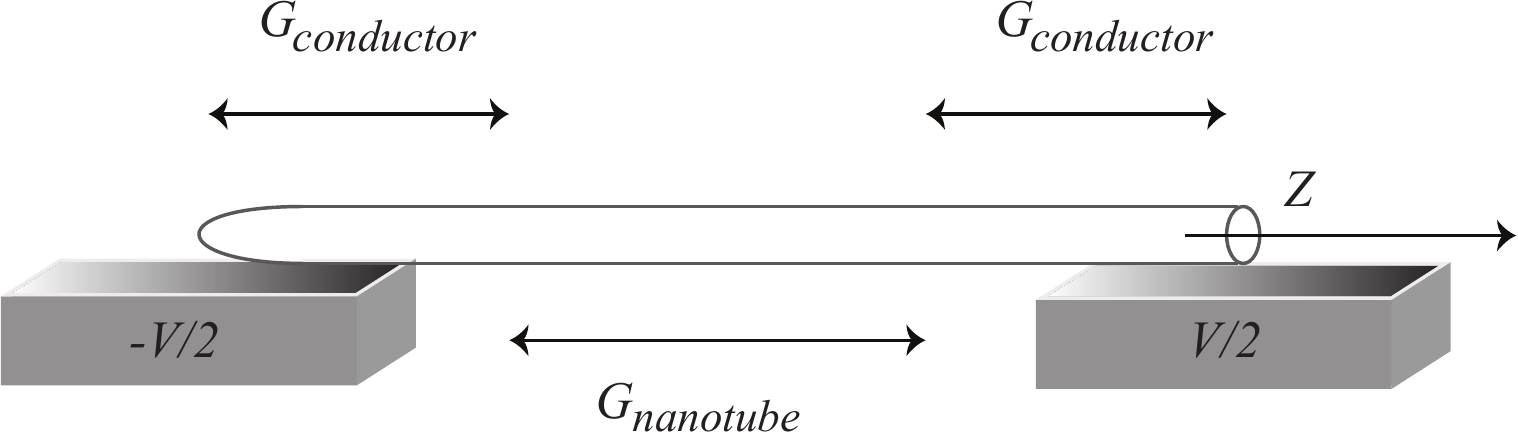}\\
  \caption{\small{(Color online)proposed mesoscopic system. SWCNT is smoothly connected two massive metallic bulk reservoirs. The conductance of this system is divided two parts: end parts and central part.}}\label{fig.dispersion}
\end{figure}

The total Hamiltonian of the system can be described by \cite{AN, TO}
\begin{equation}
H=H_0+H_1+H_{e-ph},
\end{equation}
where, the kinetic energy ($H_0$), the interaction between electrons and electrical field ($H_1$), and the e$ -  $ph coupling in the elastic regime ($H_{e-ph}$) are respectivly given by
\begin{equation}
H_0=\sum_{k} \varepsilon c^\dagger_{k} c_{k}+\sum_{q} \hbar \omega_{q} a^\dagger_{q} a_{q},
\end{equation}
\begin{equation}
H_1=\frac{ev}{2} \sum_{k} sign(\nu_z) c^\dagger_{k} c_{k},
\end{equation}
\begin{equation}
H_{int}=\sum_{k, k^\prime} M^{q}_{k, k^\prime} (a^\dagger_{-q}+ a_{q}) \, c^\dagger_{k} c_{k^\prime}.
\end{equation}
Where, $\nu_z$, is the electron velocity along the nanotube.
According to the elasticity of e-ph scattering, only the interaction of electrons with electronic field gives rise to energy loss. Then, the change in the electron current is related to the
rate of energy dissipation by  \cite{KIOOANT} :
\begin{equation}
\begin{split}
\Delta I =& \frac{dE}{dt}=\frac{d<H_1>}{dt},
\end{split}
\end{equation}
differentiating $ < H_1 > $ over the time $ t $ we obtain the
equation for $ ∆ I$, the change of the current as a result of the interaction of electron with phonons:
\begin{equation}
\begin{split}
V(\Delta I) =& \frac{1}{i \hbar}<[H_1,H_{int}]> ,
\end{split}
\end{equation}
where
\begin{equation}
\begin{split}
<O>=<Tr(\rho(t)O)> ,
\end{split}
\end{equation}
all operators are in the Dirac representation. The statistical operator $\rho(t)$ comply with equation
\begin{equation}
\begin{split}
i \hbar \frac{\partial \rho}{\partial t} =[H_{int}(t) , \,\rho(t)],
\end{split}
\end{equation}
where $\rho$ is the density operator of electrons. The change in the electronic current due to the electron-phonon coupling can be determined using the perturbation theory,$ \Delta I=\sum_{i}  \Delta I_i $, and
\begin{equation}
\begin{split}
\Delta I_0 = & \frac{1}{i \hbar \, V}  Tr \, \Big[\rho_{0} \; [H_1, \, H_{int}]\Big],
\end{split}
\end{equation}
the first order of the current variation is always equal to zero. The first non-zero term is equal to
\begin{equation}
\begin{split}
\Delta I_1 =&-\frac{1}{\hbar^2 V} \int_{-\infty}^t dt^\prime Tr \, \Big[[H, \, \rho_0]\; [H_1, \, H_{e-ph}]\Big],
\end{split}
\end{equation}
Consequently, by means of Wick's theorem, we reach the first-order correction on ballistic current, as
\begin{equation}
\begin{split}
\Delta I=\frac{-e}{\hbar^2} \sum_{k, k^\prime, q} & \Big(sign(\nu_{zk})-sign(\nu_{zk^\prime})\Big) \; \delta(\varepsilon_{k^\prime} - \varepsilon_k -\hbar \omega_q)
\\&|M^{q}_{k, \, k^\prime}|^2 \, \, \big[N_q(f_k-f_{k^\prime})+f_k(1-f_{k^\prime})\big].
\end{split}
\end{equation}
Here $\nu_z$, the parameter of sign function, shows the speed of electron and determinants the direction of electron motion. $\varepsilon_{k, k^\prime}$ refers to the electron energy, $\omega_q$ shows the phonon frequency, $M^{q}_{k,k^\prime}$ is the e$ -  $ph coupling coefficient, $N_q$ and $f_{\alpha\, \beta}$ denote Bose$ - $Einstein and Fermi$ - $Dirac statistical functions, respectively. At room temperature, the Fermi$ - $Dirac distribution function behaves as a step function \cite{ZCC}. The Fermi energy of CNT electrons is roughly $2.9$ eV \cite{PA} and the Fermi level has been considered as the reference of energy and sets to zero. Under the first order of tight$ - $binding method and Einstein approximation, correction on the current is given by 
\begin{equation}
\begin{split}
\Delta I=&\frac{-ea_0}{2\,M\,N \hbar \, t}\sum_{k_y k^\prime_y}
\\& \int_{\frac{-\pi}{\sqrt{3}a_0}}^{\frac{\pi}{\sqrt{3}a_0}} dk_z \; dk^\prime_z 
D \; \big(sign(\nu_{z k})-sign(\nu_{z k^\prime})\big)\; \delta(\frac{\varepsilon_{k^\prime} - \varepsilon_k - \hbar\omega_q}{t})
\\& \Big[\frac{1}{e^{(\beta \hbar \omega_q -1)}}\Big(\theta\,\big(\frac{eV}{2}sign(\nu_{zk}-\varepsilon_k)\big)-\theta\,\big(\frac{eV}{2}sign(\nu_{zk^\prime}-\varepsilon_{k^\prime})\big)
\\&+
\theta\,\big(\frac{eV}{2}sign(\nu_{zk})-\varepsilon_k\big)\big(1-\theta(\frac{eV}{2}sign(\nu_{zk^\prime}-\varepsilon_{k^\prime})\big)\Big)\Big],
\end{split}
\end{equation}
in which $q$ represents the branch of RBM and quantities of $k_y,\, k^\prime _y$ are taken from quantization condition for nanotube. Moreover, $D=\frac{a_0}{d}$ and $\frac{a_0}{(c_1 d+c_2 d^{-1})}$ in the first and second approximation of $\omega_{RBM}$, respectively. The first term indicates interaction with thermal phonons which change the momentum of the electron. Besides, at very low temperatures, the second term is not zero because electrons gain energy due to applying electrical field and are able to transfer from a full level to an empty one by absorbing phonon, or transfer to a lower energy level and give their energy to the lattice by emitting phonon.

\section{Results and discussion}\label{results}
\begin{figure}[!h]
  \includegraphics[width=0.47\textwidth]{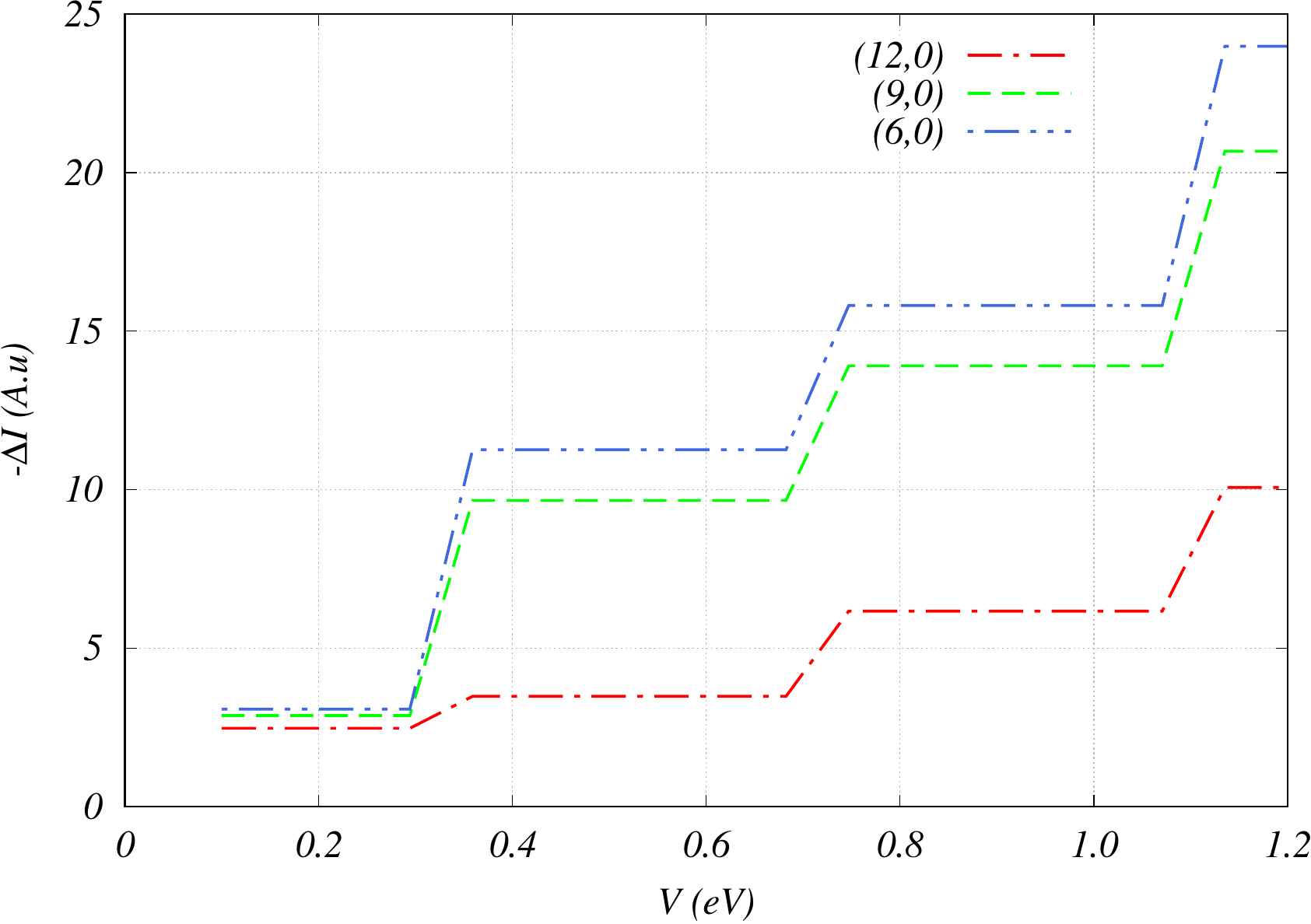}\\
  \caption{\small{(Color online) Electrical current correction diagram as a function of voltage for three different metallic zigzag CNTs by considering e-ph interaction in RBM. }}\label{fig.2}
\end{figure}

In this section, we present our results for the changes in the conductance due to the e-ph coupling, as a function of the bias voltage and temperature. Moreover, the results of different CNT diameter size and different $\omega_{RBM}$ frequencies are compared. Electrical current correction diagram of metallic zigzag CNTs as a function of Voltage has been represented in Fig.\eqref{fig.2}, for different diameters in the presence of e-ph interaction in RBM. Electrons move along nanotubes affected by the bias voltage. When the voltage increases, electrons gain sufficient energy for interaction with phonons. Interaction occurs for given quantities, because of quantization of phonon’s energy, and the electrical current increases step$ - $likely. At room temperature, high energy RBM phonons do not exist, so scattering from optical phonons contains phonon emission \cite{JSYV}.
Also step$ - $like changes of current for smaller diameter nanotubes are greater because e-ph interaction decreases by increases of nanotube diameter. In small diameter nanotubes, electrons are not able to move from bonds that do not cross the Fermi level.

\begin{figure}[!h]
  \includegraphics[width=0.47\textwidth]{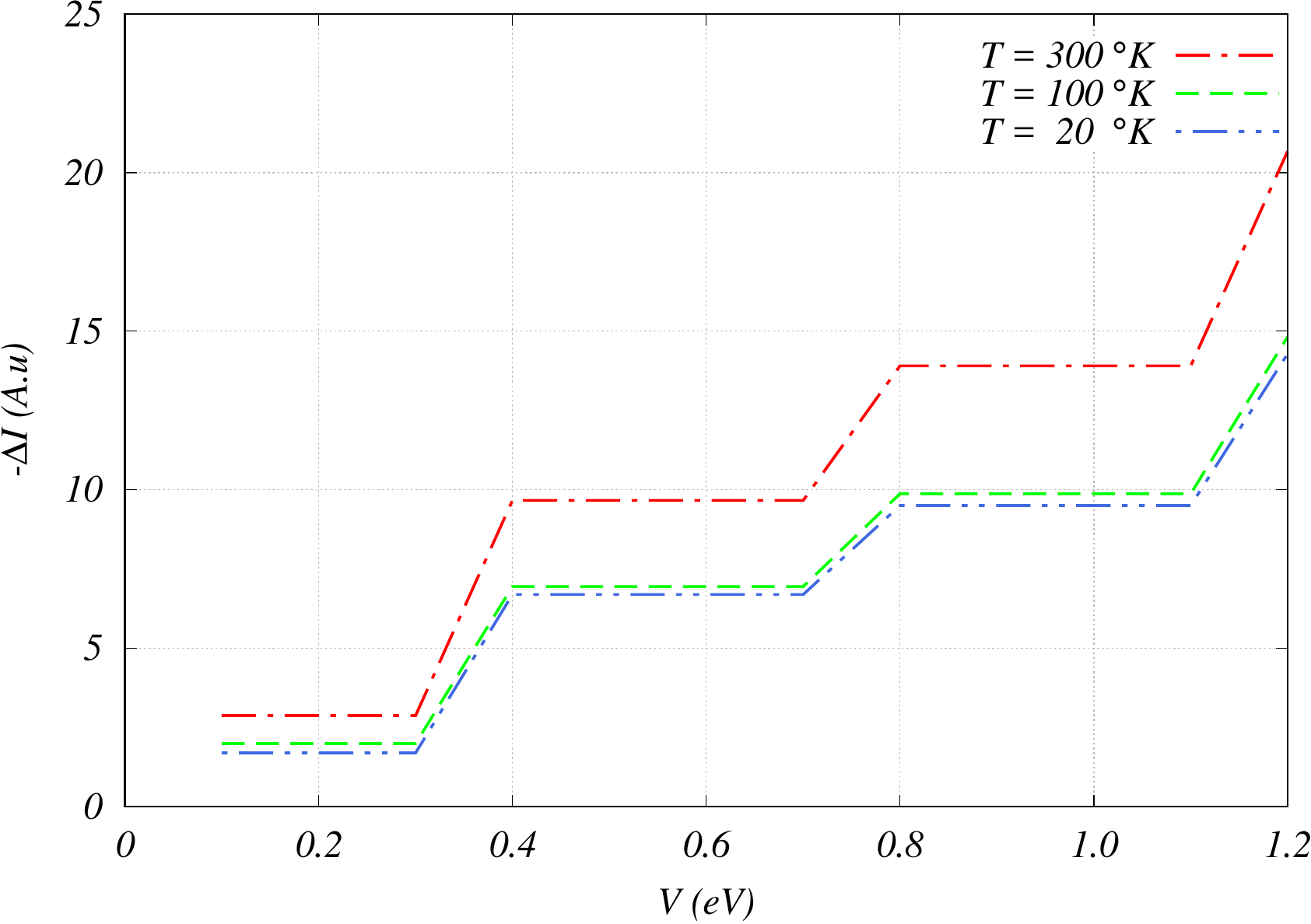}\\
  \caption{\small{(Colour online) Electrical current correction diagram as a function of voltage by considering electron-RBM phonon coupling for (9,0) zigzag CNT in three different temperatures }}\label{fig.3}
\end{figure}

Fig.\eqref{fig.3} represents diagram of current correction as a function of voltage considering electron-RBM phonon for $ (9,0) $ zigzag CNTs in three different temperatures. At temperatures that are lower than room temperature, thermal phonon’s contribution is small and excited phonons have a little effect on current changes. When temperature increases, approaching room temperature, the energy of excited modes increases and e$ -  $ph scattering has more contribution in the current decrease.

\begin{figure}[!h]
  \includegraphics[width=0.47\textwidth]{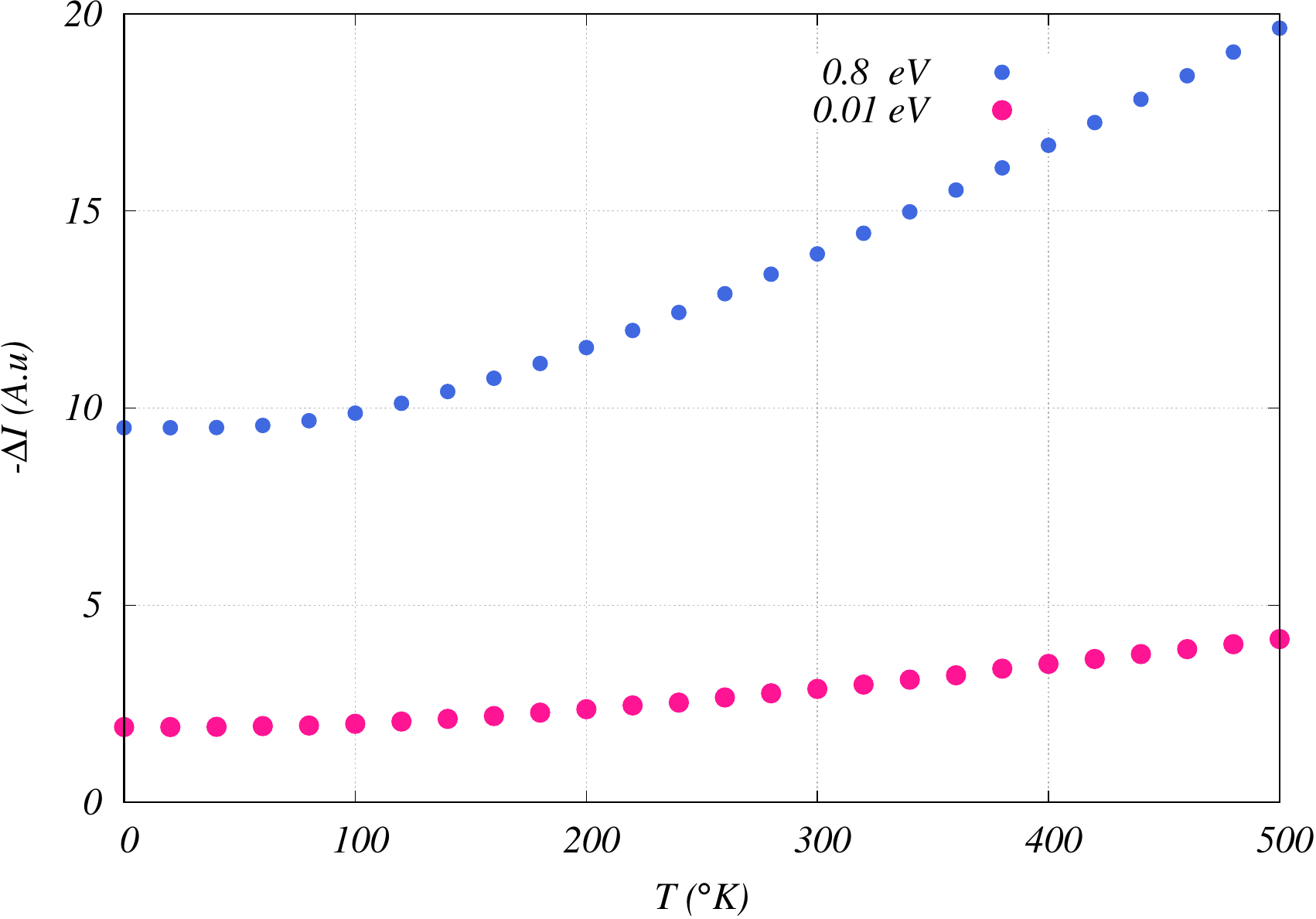}\\
  \caption{\small{(Color online) The current variation as a function of temperature, considering e-ph interaction in two different constant voltages }}\label{fig.4}
\end{figure}

Fig.\eqref{fig.4} represents the current variations as a function of temperature, considering e-ph interaction for two different constant voltages. At temperature lower than room $( < 100^\circ K )$ current can be assumed as a constant. At low temperatures and bias voltages, electrons do not have enough energy for phonon emission and optical modes cannot be excited thus e-ph interaction does not occur. But at high temperatures with excitation of radial breathing mode, current changes intensely. So by increasing the temperature, more energetic optical phonons will be excited that leads to current variations increase. As it is obvious in fig.\eqref{fig.4} at higher voltages because of the presence of energetic electrons and high energy phonons, current variations have more increase.
 
\begin{figure}[!h]
  \includegraphics[width=0.4\textwidth]{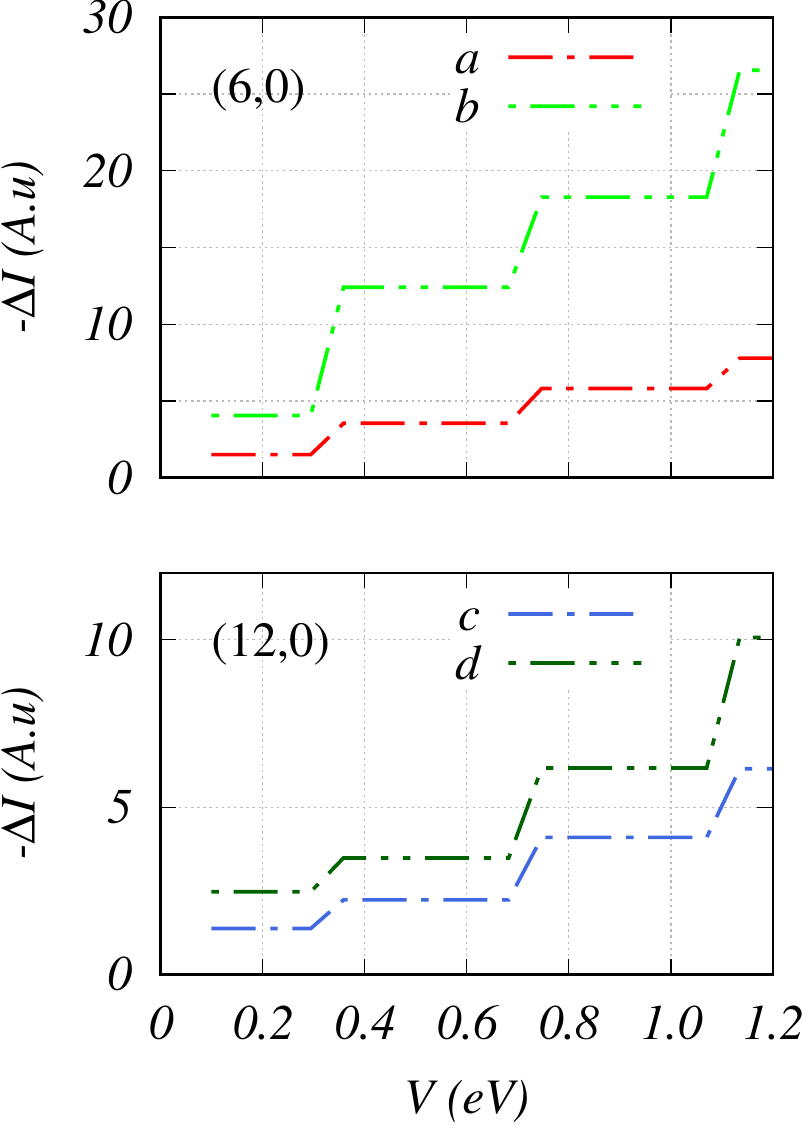}\\
  \caption{\small{(Color online) A comparison between two different approximations has been made for two different nanotube diameter sizes. The top panel belongs to $ (6,0) $ nanotube while the bottom one represents the result of $ (12,0) $. Here, the curves (a,c) and (b,d) are achieved based on the second and first approximation $\omega_{RBM}$ respectively.}}\label{fig.fig562p}
\end{figure} 

The second approximation has been used for $\omega_{RBM}$ due to considering distortions of the band to calculate the variations of the current and has compared with results of first approximation for $\omega_{RBM}$ in Fig.\eqref{fig.fig562p} (a) and (b). This figure represents variations of the current versus the bias voltage considering electron-RBM phonon interaction in (6,0) zigzag nanotube by second and first approximations for $\omega_{RBM}$ and  Fig.\eqref{fig.fig562p} (c) and (d) represents variations of current versus bias voltage considering electron-RBM phonon interaction in (12,0) zigzag nanotube by second and first approximations for $\omega_{RBM}$. It is observed that current variations of nanotube considering e-ph interaction in Fig.\eqref{fig.fig562p} (a) and (b) are more noticeable than Fig.\eqref{fig.fig562p} (c) and (d). Indeed, considering more effects of nanotube diameter, current variations decrease. Based on a comparison of Fig.\eqref{fig.fig562p} (a) and (b) with (c) and (d), it has been found that by considering more effects of nanotube diameter, reduction of current variations in small diameter nanotubes, is noticeably more than large diameter nanotubes. In other words, the decrease in the difference between results of two approximations is related to decrease of e-ph interaction in larger diameter nanotubes which causes effects of nanotubes diameter to weaken in current variations.

\section{Conclusion}\label{conclusion}
In conclusion, we demonstrated that, as the bias voltage increases sufficiently, the creation of a phonon gives rise to specific current variations which directly depends on the diameter. Current variations, based on the electron-RBM phonon coupling, are step$ - $like function of bias voltage. Besides, one can observe that current variations increase by decreasing the diameter of nanotube since electron-RBM phonon coupling reduces as the diameter of nanotube rises because in this case CNTs behave as same as graphene sheet and the electron-RBM phonon interaction can be neglected. Moreover, as infinite CNTs has the continuous energy spectrum along its axis direction, when the energy is inducted into the system, at the first, low energy phonons are created and following the continuous energy increase, higher energy phonons emerge. To shed light on the study of CNTs properties, we proposed a simple method for creation and characterization of RBM phonons of metallic zigzag nanotubes by electron injection. As in experimental realization, a branch of CNTs with different shapes and sizes are produced, it is essential to characterize their diameter. Since RBM mode determines the geometry and structure of CNT, by following the peaks of the conductance we can trace the different CNTs in a sample. So this approach can be a proper and simple method for the identification of CNTs.

\section*{Acknowledgement}
R. T. acknowledges Fariborz Parhizgar for his fruitful discussion and support and also thanks Saeed Amiri and Mahroo Shiranzaei for their support during preparing the last parts of this paper.


\end{document}